\UseRawInputEncoding
\documentclass[preprintnumbers, prl, floatfix,twocolumn,
preprintnumbers, letterpaper, superscriptaddress]{revtex4}
\usepackage{amsfonts}
\usepackage{amsmath}
\usepackage{amssymb,epsf}
\usepackage{latexsym}
\usepackage{graphicx,epsfig}
\usepackage{amssymb}
\usepackage{subfigure}
\usepackage[colorlinks=true,citecolor=blue,linkcolor=blue,urlcolor=black]{hyperref}
\usepackage{epstopdf}
\usepackage{color}

\def\be{\begin{equation}}
\def\ee{\end{equation}}
\def\ba{\begin{eqnarray}}
\def\ea{\end{eqnarray}}
\def\la{\langle}
\def\ra{\rangle}

\begin{document}

\title{Topological line in frustrated Toric code models}
\author{M. H. Zarei}
\email{mzarei92@shirazu.ac.ir}
\affiliation{Department of Physics, School of Science, Shiraz University, Shiraz 71454, Iran}
\author{J. Abouie}
\email{jahan@iasbs.ac.ir}
\affiliation{Department of Physics, Institute for Advanced Studies in Basic Sciences (IASBS), Zanjan 45137-66731, Iran}

\begin{abstract}
Typical topological systems undergo a topological phase transition in the presence of a strong enough perturbation. In this paper, we propose an adjustable frustrated Toric code with a "topological line" at which no phase transition happens in the system and the topological order is robust against a non-linear perturbation of arbitrary strength. This important result is a consequence of the interplay between frustration and nonlinearity in our system, which also causes to the emergence of other interesting phenomena such as reentrant topological phases and survival of the topological order under local projection operations. Our study opens a new window towards more robust topological quantum codes which are cornerstones of large-scale quantum computing.
\end{abstract}
\maketitle

\section{Introduction}\label{s1}

Topological phases are states of matters that are robust against local perturbations \cite{wen1, wen2, t1, t2, t3, 20, bravyi}. Designing topological systems is of crucial importance in a wide range of practical applications from quantum computing \cite{del,18,19,200,22,24} and topological spintronics \cite{Fan,Beenakker} to soft matter and mechanical systems \cite{Nagel, Ma19}, and searching for systems with rich topological properties is one of the growing interests in both theoretical and experimental physics  \cite{Hasan,Haldane,Armitage,Ozawa,zarei2016,Ahmadi2020}.
Toric codes (TCs) are a kind of topological quantum models, characterizing by their robust topological degeneracy.
They were first introduced for topological quantum memory \cite{Kitaev2003,15}, and have gained significant importance in recent years in the context of large-scale quantum computing \cite{pra,thr,sur,natu}. 
Many attentions have been devoted to the investigation of the
effects of different types of perturbations including external magnetic fields \cite{robustness,ala,schmit,29,wu,sch,jadid}, Ising interactions \cite{karimipour,robustness2} and local tensor perturbations \cite{nor} on the topological properties of TCs on different lattices. 

One of the important challenges is to find different properties that lead to more robustness of the topological orders. Recently, the interplay of topology and frustration has been studied and demonstrated that geometrical frustration leads to the further robustness of the TC state \cite{frus1, frus2}. Furthermore, it has also been shown that the perturbations that couples nonlinearly to a topological model can lead to further robustness \cite{castel}.  Nevertheless, it is believed that regardless of the kind of local perturbation, topological orders will be eventually destroyed at a point where a topological-trivial phase transition occurs in the system. Hence, it is raising a question of whether it is possible to construct an adjustable system with an everlasting topological order.

\begin{figure}[h]
	\centering
	\includegraphics[width=7cm,height=5cm,angle=0]{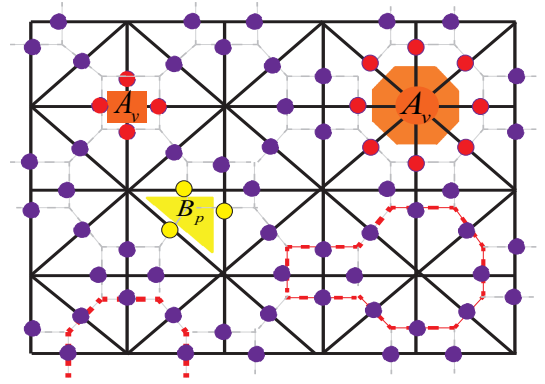}
	\caption{(Color online) An illustration of the TC on the UJ lattice (black solid lines). The red and yellow qubits correspond to the vertex and plaquette operators, respectively. The dual square-octagonal lattice is displayed by the gray dashed lines where each vertex of the TC can be represented by a simple loop in the dual lattice. The red dashed lines show a loop configuration on the dual lattice. The qubits belonging to these loops are in the state $|1\ra$, an eigenstate of $Z$ with eigenvalue $-1$.} \label{Fig:UJ}
\end{figure}

In this paper, we examine the combined effects of frustration and nonlinearity by proposing frustrated quantum models composed of a TC coupled nonlinearly to an external perturbation, and show that their ground states on various lattices possess a topological phase with interesting properties, not seen in other topological systems so far. We demonstrate that there is a topological line (TL) at which (i) by tuning the physical parameters of the system it remains in the topological phase in the presence of perturbation of any strengths, (ii) the topological order of the ground state survives under local projection operations. In order to identify this TL, and to figure out the physical mechanism behind such a non-trivial phenomenon, we define a topological string parameter and show that the TL is a line separating two topological regions with different behaviors of the string parameter. Our frustrated TC (FTC) models have also another interesting feature, in which a reentrant topological phase transition occurs in the ground state phase diagram of the system while the perturbation parameter increases.
This phenomenon signifies the reversibility of the topological order to the system in the presence of strong perturbations. 

This paper is structured as follows: In Sec.(\ref{s2}), we introduce our FTC model and find its phase diagram by using a mapping to the frustrated Ising model. Then in Sec.(\ref{s3}), we analyze the phase diagram to find that there is a reentrant topological phase transition. More importantly, in Sec.(\ref{s4}) we introduce the topological line which is charactrized by a topological string parameter and accordingly we explain how the system is robust against the perturbation along the topological line. Finally in Sec.(\ref{s5}), we show that the system is also robust against local projections at the topological line.

\section{Toric code in presence of a non-linear perturbation}\label{s2}

A TC model is described by the Hamiltonian:
 \begin{equation}
H_{\rm TC}=-\sum_p B_p  -\sum_v A_v,
\end{equation}
where $B_p= \prod_{i\in \partial p} Z_i$ and $A_v=\prod_{i\in v}X_i$ are respectively the plaquette and vertex operators with $Z$ and $X$ being the Pauli operators, $i\in \partial p$ refers to the qubits around the plaquette $p$, and $i \in v$ refers to the qubits incoming to the vertex $v$ (see Fig. \ref{Fig:UJ}, the yellow and red dots). This Hamiltonian is exactly solvable and its ground state is given by
$|\psi\ra = \prod_{v}(I+A_v) |0\ra^{\otimes N}$,
where $|0\ra$ is an eigenstate of $Z$ with eigenvalue 1, $I$ refers to the identity operator and $N$ is the number of qubits. Since each vertex operator can also be represented by a loop operator in a dual lattice (in Fig. \ref{Fig:UJ}, we have illustrated the dual of a Union-Jack (UJ) lattice by the gray dashed lines), the ground state $|\psi\rangle$ is also a loop-condensed state (each loop is made of $|1\ra$ states, and a loop-condensed state is a superposition of all loop-configurations in a sea of $|0\ra$ states).
It is important to note that, with periodic boundary condition the initial lattice is attached to a torus with non-trivial loops which lead to different degenerate ground states. Since these degenerate states correspond to different topological classes, they are robust against local perturbations.

Now, let us introduce a perturbation to the system via the Hamiltonian:
\begin{equation}\label{eq2}
H_e=\sum_v e^{-\beta \sum_{i\in v}J_i Z_i},
\end{equation}
where $\beta$ and $ J_i$ are parameters controlling the order of the system. Using a Taylor expansion, one can see that $J_i$ can be served as the magnetic moment of the qubit $i$, adjustable within the system, and $\beta$ refers to an inevitable perturbation, arising from an effective field coupled to the moments.
At small values of $\beta$, the above perturbation reduces to a Zeeman term where the field is coupled linearly to the spins, however for larger values of $\beta$, nonlinear effects arising from the local multi-spin interactions around each vertex (spins that are the nearest neighbor of vertices) become important in the topological characteristics of the system, especially in the robustness of the topological order \cite{expan}.
In practice, the perturbation $\beta$ is unavoidable, and the topological order of the system is finally lost by strong perturbations, however our FTC system has an additional ability that by adjusting the moments of the qubits ($J_i$), one can achieve a topological order, robust against perturbation of any strengths. The systems with adjustable magnetic moments and multi-body interactions can be realized in experiment with current technologies by cold atoms (to find how multi-body interactions can be implemented, see the protocol recently proposed in Ref. \cite{Bohrdt} and references therein).
Here, we set $J_i$ to $J_1$ for all qubits on the vertical and horizontal edges and $J_2$ for all on the diagonal edges.

The frustrated Hamiltonian, $H_{\rm FTC}=H_{\rm TC}+H_e$, is indeed a type of stochastic matrix form Hamiltonians \cite{castel2005}, and its ground state can be exactly found \cite{castel,zarei20,zarei18} as:
\begin{equation}\label{eq3}
|G(\beta, \{J_i \})\ra =\frac{1}{\sqrt{\mathcal{Z}(\beta)}}
e^{\frac{\beta}{2}\sum_i J_i Z_i}|\psi \ra,
\end{equation}
where $|\psi\ra$ is the ground state of the Hamiltonian $H_{\rm TC}$. By applying the operator $e^{\frac{\beta}{2}\sum_i J_i Z_i}$ on the state $|\psi\ra$, we will have a superposition of loop-configurations with amplitudes $e^{\frac{\beta}{2}\sum_i J_i \sigma_i}$, where $\sigma_i =-1(+1)$ for links with qubits in the state $|1\ra$ ($|0\ra$).
The normalization factor in Eq. (\ref{eq3}) is thus obtained as $\mathcal{Z(\beta)}=\sum_{lc}e^{\beta \sum_i J_i \sigma_i}$, where the summation runs over all loop-configurations. This function is nothing but the partition function of a classical Ising model on the UJ lattice, where the parameter $\beta$ plays the role of the inverse of the thermal energy $k_B T$, $J_i$ is the local exchange interaction between the two nearest neighbor Ising spins, $S_i$ and $S_{i+1}$, located at the UJ lattice points $i$ and $i+1$, and $\sigma_i=S_iS_{i+1}$ is equal to +1 (-1) when the nearest neighbor spins are parallel (antiparallel) \cite{Steph1}.
In the low temperature expansion of this partition function, we can see that each spin-configuration is also represented by a loop-configuration in the dual square-octagonal lattice.

\begin{figure}[t]
	\centering
	\includegraphics[width=8cm,height=6cm,angle=0]{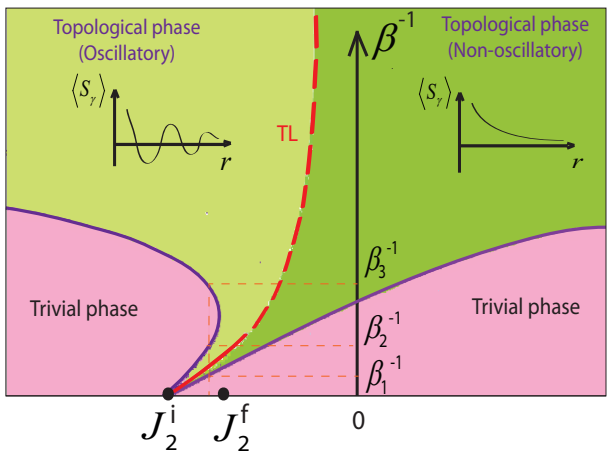}
	\caption{(Color online) The ground state phase diagram of our FTC on the UJ lattice. The vertical and horizontal axes are respectively $\beta^{-1}$ and $J_2$ (we set $J_1=1$). The green and the pink regions are respectively the topological and the trivial phases which are seperated by the phase boundaries with equations: $\cosh (4\beta J_1)=e^{-4\beta J_2}\pm 2e^{-2\beta J_2 }$, where the signs $+$ and $-$ are respectively for the blue curves at the right and the left sides of the red dashed line. The dashed red line is the TL. This line divides the green region into two parts where the topological string parameter decays oscillatory and non-oscillatory with respect to the string length. In the interval $J_2^i<J_2<J_2^f$ a reentrant topological phase appears at small values of $\beta^{-1}$ (the green region between $\beta_1^{-1}$ and $\beta_2^{-1}$). For the case of $J_1=1$, the parameters $J_2 ^i$ and $J_2 ^f$ are approximately equal to $-1$ and $-0.9$.} \label{Fig:phase-diagram}
\end{figure}

It is intuitive to compare the ground state of the TC on a square lattice in the presence of a uniform magnetic field with the ground state of our FTC model. 
In both cases they are a superposition of loop-configurations, but the effects of perturbations are different. 
In the former case, the perturbation causes the generation of open strings \cite{robustness, frus1}, while in ours, owing to nonlinearities arising from the local multi-body interactions, the perturbation only changes the amplitudes of the loop-configurations. For example, in the case of $J_1=J_2 =+1$, the amplitudes in Eq. (\ref{eq3}) are in the form of $\sim e^{-\beta l}$, where $l$ is the total perimeter of all loops. This term shows that the perturbation acts as a "{\it tension}", and an increase of $\beta$ decreases the amplitude of large loops. In particular, in the limit of $\beta \rightarrow \infty$, the amplitude of all loop-configurations goes to zero and the final state will be the product state $|00...0\ra$. Since the initial state at $\beta =0$ is the topological loop-condensed state $|\psi\ra$, it is concluded that there must be occurred a topological-trivial phase transition by the increase of $\beta$ from zero to $\infty$ (more details for the ground state phase diagram of the TC on a simple unfrustrated square lattice has been addressed in \cite{castel,zarei20}).
On the other hand, the situation is different if $J_1=J_2=-1$. In this case, the amplitudes in Eq. (\ref{eq3}) are simplified as $\sim e^{\beta l}$. In contrast to the previous case, here the perturbation $e^{\beta l}$ plays the role of a "{\it pressure}", and an increase of $\beta$ leads to the generation of loops with larger total perimeter. In the limit of $\beta\rightarrow\infty$, the loop-configuration with maximum total perimeter is dominant and the ground state is the product state $|11...1\ra$. 
Finally, in the case of $J_1 \geq 0$ and $J_2 \leq 0$, both the string tension and pressure are present in the system. In this case
the interplay of the string tension and pressure causes the system to be frustrated. 
The simultaneous presence of the frustration and nonlinearities arising from the local multi-body interactions is in the favor of topological robustness, and leads to  
the interesting phenomena discussed in the following sections. 

\section{Reentrant topological phase} \label{s3}
In order to obtain the topological phase transition points in our FTC model, we investigate the behavior of the ground state fidelity; $F=\la G(\beta, \{J_1 , J_2 \})|G(\beta +d\beta , \{J_1 , J_2\})\ra$.
Employing a Taylor expansion, the ground state fidelity \cite{zanardi} is readily obtained in terms of the specific heat ($C_V$) of the Ising model as:
\begin{equation}\label{fid}
F \simeq1-\frac{C_V}{8 \beta^2}d \beta ^2,
\end{equation}
where we have used the equality
$C_V=\beta^2\frac{\partial ^2 \ln\mathcal{Z}}{\partial \beta^2}$.
The above relation indicates that corresponding to a phase transition temperature in the classical model where the specific heat shows a singularity, there must be a topological phase transition point, where the ground state fidelity becomes singular.

Fortunately, the phase diagram of the Ising model on different lattices such as the UJ and the triangular lattices is exactly known \cite{Steph1,diep}. Here, as an interlude we explain the classical phase diagram of the Ising model on the UJ lattice with $J_1>0$ and $J_2$ couplings. This model possesses two different phases with ferromagnetic and antiferromagnetic long range orders at low temperatures and a paramagnetic disorder phase at high temperatures. For large positive (negative) values of $J_2/J_1$, strong ferromagnetic (antiferromagnetic) couplings cause the system to be in the ferromagnetic (aniferromagnetic) phase at low temperatures. However, at intermediate strengths of $J_2/J_1<0$, the system behaves exotically, it is in the paramagnetic phase at high temperatures and goes to the antiferromagnetic phase by decreasing temperature. By further decreasing of temperature, we expect the staggered magnetization to increase and the true long range Neel order to form in the classical phase diagram, but this will not happen and amazingly a phase transition occurs to the paramagnetic phase. This phenomena which is a consequences of frustration is called reentrant phase transition. 

Now we come back to our frustrated TC model on the UJ lattice, where $\beta$ is the perturbation parameter.  Since $\beta^{-1}$ plays the role of temperature in the classical Ising model, the paramagnetic phase corresponds to a topological order at small values of $\beta$, and the ordered ferromagnetic and antiferromagnetic phases at low temperatures correspond to a topologically trivial phase at large values of $\beta$.  In Fig. \ref{Fig:phase-diagram}, we have presented the ground state phase diagram of our FTC model on the UJ lattice.
Let us focus on the interesting region of $J_2 ^{i}< J_2< J_2 ^{f}$.
At large values of $\beta^{-1}$ (or small strengths of perturbation) the ground state 
possesses a topological order, however this order can not persist in the presence of stronger perturbations (or smaller $\beta^{-1}$) and disappears eventually at a transition point ($\beta_3^{-1}$) where the system enters to a trivial phase. Now, we expect the system to be locked in the trivial phase, and perturbations wash out the topological order completely, however we see that amazingly a " reentrant topological phase transition " occurs in the system and the topological order revives at the second phase transition point ($\beta_2^{-1}$).
We have also examined our FTC model on other geometrically frustrated lattices such as Kagome and triangular lattices (the results are not shown here), and observed the above mentioned phenomenon also in these systems. Actually, the emergence of a reentrant topological phase is a dramatic impact of frustration and nonlinearities arising from the multi-body interactions introduced by the Hamiltonian in Eq. (\ref{eq2}) to the system.

\section{Topological line}\label{s4}
As we discussed, there is a topological order in the ground state phase diagram of our FTC model which is robust against small perturbations, $\beta$, but 
disappears in the presence of strong perturbations when a topological-trivial phase transition occurs in the system.
However, surprisingly we see that exactly at a line in the topological phase, the system remains in the topological phase in the presence of any strength of the parameter $\beta$. We call this line as " Topological line (TL)", because the topological order survives at this line regardless of the strength of $\beta$ (see the dashed line in Fig. \ref{Fig:phase-diagram}).
\begin{figure}[t]
\centering
\includegraphics[width=6cm,height=4cm,angle=0]{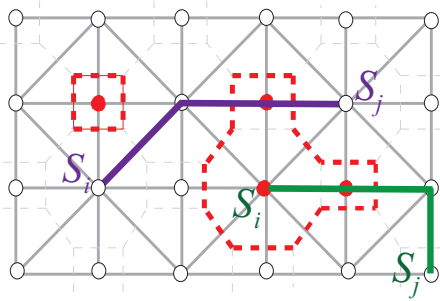}
\caption{(Color online) A UJ lattice with Ising spins (white dots). The solid lines are strings that connect the two spins $S_i$ and $S_j$ located at the lattice points $i$ and $j$. Red and white dotes are respectively spin $+1$ and $-1$. The blue (green) string crosses a loop for even (odd) times. } \label{fig6}
\end{figure}
In order to obtain the equation of this TL in the ground state phase diagram, we utilize the TC-Ising mapping explained in the previous section. 
In the language of the classical Ising model on the UJ lattice, the TL is a disorder line in the paramagnetic phase which separates two regions with different two-point correlation functions. In the region next to the ferromagnetic phase the two-point correlations decay nonoscillatory by increasing the separation distance of the Ising spins, while oscillatory in the region next to the antiferromagnetic phase.
The two-point correlation functions in the Ising model are equal to the signed summations of the Boltzmann weights where the sign behind each Boltzmann weight is determined by the sign of $S_i S_j$ ($S_i$ is an Ising spin located at UJ lattice point $i$). Since, each spin-configuration on the UJ lattice is represented by a loop-configuration on its dual lattice, we can express the two-point correlation functions as signed summations of the Boltzmann weights corresponding to different loop-configurations. In order to determine the sign of the mentioned Boltzmann weights, we pull a string $\gamma$ between the two spins $S_i$ and $S_j$ on the UJ lattice (see Fig. \ref{fig6}). 
For spin-configurations with parallel (antiparallel) $S_i$ and $S_j$, the corresponding loop-configurations on the dual lattice will cross the string $\gamma$ for even (odd) times (see Fig. \ref{fig6}), and the sign of the Boltzmann weight becomes $+1~(-1)$.
Now, we define a string operator as $\mathcal{S}_\gamma =\prod_{i \in \gamma} Z_i$ where $i\in \gamma$ refers to all qubits belonging to the string $\gamma$ \cite{string}. 
The expectation value of this operator in the ground state of our FTC model is equal to the two-point correlation function $\la S_i S_j \ra$ in the Ising model, i.e.
$\la S_i S_j \ra =\la G(\beta)|\mathcal{S}_\gamma |G(\beta)\ra$.
By using this relation, we can obtain the equation of the TL as:
\begin{equation}\label{tl}
\cosh(4\beta J_1)=\exp(-4\beta J_2).
\end{equation}
This TL divides the topological phase into two parts as shown in Fig. \ref{Fig:phase-diagram}. In these regions the topological string parameter decays oscillatory and nonoscillatory by increasing the length of the string, respectively. These different behaviors can be well interpreted by comparing the strengths of the string tension and pressure in the  FTC model. Actually, in the topological phase at the left of the TL, the pressure
causes several small loops to be generated in the system, and
hence loop-configurations with larger total perimeters play the dominant role in determining the ground state of the system.
It should be noted that loop-configurations with large perimeters are those which are generated from several small loops.
Accordingly, a typical string $\gamma$ with the length of $r$ successively crosses the small loops, and consequently the sign of the string parameter $\la\mathcal{S}_\gamma(r)\ra$ oscillates by increasing $r$.
On the other hand, in the topological phase at the right of the TL, the string tension plays the dominant role, and the loop-configurations with very few small loops are crucial in determining the ground state. In this region, unlike the topological phase at the left of the TL, the sign of the string parameter does not change by $r$ (see the inset plots in Fig. \ref{Fig:phase-diagram}). Finally, at the TL, the reciprocal effects of tension and pressure are balanced, resulting in the formation of closed loops with different sizes. The stability of various loops leads to robustness of the topological order at the TL.

\section{Topological line and robustness against local projections}\label{s5}
\begin{figure}[t]
\centering
\includegraphics[width=5cm,height=4cm,angle=0]{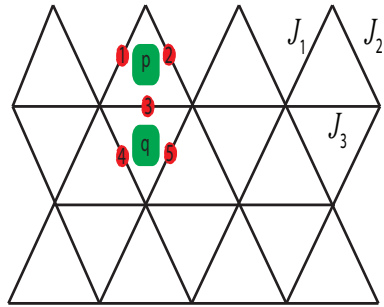}
\caption{(Color online) Schematic illustration of our FTC model on a triangular lattice with $J_1$, $J_2$, and $J_3$ tuning parameters. The plaquettes $p$ and $q$ share the qubit 3. At the TL, by projecting out the red qubits denoted by 1, 2, and 3 by the state $\la +|$, leads to the same frustrated toric code model on a reduced triangular lattice where the plaquette $p$ is removed.} \label{fig7}
\end{figure}

The TL has also another characteristic at which local projection operations are not able to destroy the topological order of the ground state. To explain this phenomenon we start with an important property of the TC state $|G (\beta =0)\ra=|\psi \ra$, in which if we apply a projection operator like $|+\ra \la +|$, with $|+\ra$ being an eigenstate of the Pauli operator $X$, on a single qubit of the state $|\psi\ra$, it removes the corresponding edge from the lattice and the quantum state of the rest will be again a TC state \cite{ras,zar}.  
Now, we consider our FTC model on a triangular lattice with three tuning parameter $J_1$, $J_2$, and $J_3$ corresponding to three qubits of each triangle, see Fig. 4. If we project out the three spins of a triangle, the quantum state of the rest is given by:
\begin{equation}\label{qw}
|\tilde{G}(\beta)\ra=\la +_1+_2+_3|G(\beta)\ra,
\end{equation}
where $|G(\beta)\ra$ is the ground state of the FTC on a triangular lattice. 
By applying the mentioned projection operators sequentially on the ground state, different excitations will be created in the system. They move in the lattice and destroy the topological order of the ground state.
In general, the reduced ground state $|\tilde{G}(\beta)\ra$ is not the same as the initial state $|G(\beta)\ra$. However, at the TL they are exactly identical. In order to prove this important property of the TL, let us project out the three spins of a triangle, for example the triangle denoted by $p$ shown in Fig. \ref{fig7}.  The ground state of our FTC is written as $|G(\beta)\ra =\frac{1}{\mathcal{Z}}e^{\sum_i \frac{K_i}{2} Z_i}|\psi\ra$ where $K_i =\beta J_i$ and $|\psi\ra$ is the pure TC state on the triangular lattice which can be written in the form of $\prod_{\Delta}(1+B_{\Delta})|+\ra^{N}$, where $\Delta$ refers to triangular plaquettes, $N$ is the number of edges and $B_{\Delta}=\prod_{i\in \Delta}Z_i$ is plaquette operator. Substituting $|G(\beta)\ra$ in Eq. (\ref{qw}), we obtain
\begin{eqnarray}
\nonumber&&|\tilde{G}(\beta)\ra=e^{\sum{i\neq1,2,3}\frac{K_i}{2} Z_i}\prod_{\Delta \neq p,q}(1+B_\Delta )|+\ra^{\otimes (N-3)}\times\\
\nonumber&&\la+_1+_2+_3|e^{\sum_{i=1}^3\frac{K_iZ_i}{2}}(1+B_p)(1+B_q)|+_1+_2+_3\ra,\label{}
\end{eqnarray}
where $B_p =Z_1 Z_2 Z_3$ and $B_q =Z_3 Z_4 Z_5$. Here, $p$ and $q$ are the  label of triangles which share the qubit 3 (see Fig. \ref{fig7}). 
By doing straightforward calculations we achieve the ground state of the FTC as:
\begin{eqnarray}
\nonumber&&|\tilde{G}(\beta)\ra=e^{\sum{i\neq1,2,3}\frac{K_i}{2} Z_i}\prod_{\Delta \neq p,q}(1+B_\Delta )|+\ra^{\otimes (N-3)}\times\\
&&[A(K_1, K_2 ,K_3)+ B(K_1, K_2, K_3)Z_1 Z_2],
\label{qw0}
\end{eqnarray}
with
\begin{eqnarray}
&& \nonumber A(K_1, K_2, K_3)=\prod_{i=1}^3\cosh(K_i/2)+\prod_{i=1}^{3}\sinh(K_i/2),\\
&& \nonumber B(K_1, K_2, K_3)=\prod_{i=1}^2\cosh(K_i/2)\sinh(K_3/2)\\
&&~~~~~+\prod_{i=1}^{2}\sinh(K_i/2)\cosh(K_3/2).
\end{eqnarray}
Now, if we set $B(K_1 ,K_2 , K_3)$ to zero, according to Eq. (7), the ground state $|\tilde G(\beta)\rangle $ is identical to $|G(\beta)\rangle $ on the reduced triangular lattice where the triangle $p$ is removed. Consequently, for those $K_1$, $K_2$ and $K_3$ which satisfy the constraint $B(K_1 , K_2 , K_3)=0$, the final state will be again the FTC state. This constraint is nothing but the equation of the TL. Finally, we point out that regarding to the classical-quantum mapping, the above process for our FTC model is similar to the " dimensional reduction " in frustrated Ising model where one can obtain the partition function of the classical Ising model at the disorder line by tracing over spin degrees of freedom row-by-row \cite{diep}.

The invariance of our FTC ground state under the local projections implies that independent of the strength of perturbations, the excitations are suppressed at the TL by frustration and nonlinearity, and the topological order survives under the above local projections. The existence of such a TL in our FTC models is crucial in practical applications. In particular, one can imagine that in the presence of a perturbation with arbitrary strength, we can tune the moment $J_2$ related to physical qubits living in the diagonal edges of the lattice so that the system remains permanently in the topological phase. Actually, having two types of physical qubit with different moments, $J_1$ and $J_2$, is an additional ability of our FTC which is absent in the TC. 

\section{Summary and outlook}
The robustness of topological orders against local perturbations is of crucially importance for modern practical applications, and constructing highly-robust topological systems is one of the most important current challenges. In this paper we have taken an important step forward by introducing realizable models with a robust topological order. We demonstrated that the interplay of frustration and nonlinearity in our system leads to the formation of a TL at which no phase transition occurs in the system, and the topological order is robust even against local projection operations. We also found another interesting phenomena, not seen in other frustrated systems, that the ground state of our frustrated Toric code model experiences a reentrant topological phase transition. This phenomenon signifies the reversibility of the topological order to the system in the presence of frustration.

Our paper paves the way for future studies for implementing more robust topological quantum codes. In particular, it is interesting to explore the existence of such a TL in different quantum codes including color codes and fracton codes, employing proper quantum-classical mappings.

\section*{Acknowledgement}
The authors would like to thank S. S. Jahromi, A. Ramezanpour, A. Montakhab and L. Memarzadeh for fruitful discussions.


\begin{thebibliography}{99}
\bibitem{expan}
By expanding the exponential in Eq. (\ref{eq2}) with respect to $\beta$, we have
$\sim 1-\beta J_iZ_i+\beta^2J_iJ_jZ_iZ_j+\beta^3 J_iJ_jJ_kZ_iZ_jZ_k+\dots$, where there is a summation on $i, j, k$.
\end{thebibliography}
\end{document}